\documentclass[english]{article}
\usepackage[T1]{fontenc}
\usepackage[latin1]{inputenc}
\usepackage{geometry}
\usepackage{float}
\usepackage{graphicx}
\usepackage{setspace}
\makeatletter

\usepackage{amsmath,amssymb}
\date{}

\usepackage{babel}
\makeatother
\begin{document}
\begin{center}{\LARGE Dynamical Symmetry Breaking with Vector Bosons }\end{center}{\LARGE \par}
\smallskip{}

\begin{onehalfspace}
\begin{center}{\large G. Cynolter$^{*}$, E. Lendvai$^{*}$ and G.
Pócsik$^{\dagger}$ }\end{center}{\large \par}
\smallskip{}

\begin{center}\textit{$^{*}$Theoretical Physics Research Group of
Hungarian Academy of Sciences, Eötvös University, Budapest, 1117 Pázmány
Péter sétány 1/A, Hungary}\\
\textit{$^{\dagger}$Institute for Theoretical Physics, Eötvös Lorand
University, Budapest, 1117 Pázmány Péter sétány 1/A, Hungary }\end{center}
\smallskip{}
\end{onehalfspace}

\begin{abstract}
An alternative nonrenormalizable low energy effective model of electroweak
symmetry breaking is proposed. In the standard model of electroweak
interactions the Higgs doublet is replaced by a complex vector doublet
and a real vector singlet. The gauge symmetry is broken dynamically
by a mixed condensate of the doublet and singlet vector fields. Gauge
fields get their usual standard model masses by condensation. The
new vector matter fields become massive by their gauge invariant selfcouplings
and expected to have masses of few hundred GeV. Fermions are assigned
to the gauge group in the usual manner. Fermion masses are coming
from a gauge invariant fermion-vector field interaction by a mixed
condensat, the Kobayashi-Maskawa description is unchanged. Perturbative
unitarity estimates show that the model is valid up to 2-3 TeV. It
is shown that from the new matter fields a large number of spin-one
particle pairs is expected at future high energy $e^{+}e^{-}$ linear
colliders of 500-1500 GeV. The inclusive production cross section
of new particle pairs is presented for hadron colliders, while at
the Tevatron the new particle production is too low, at the LHC the
yield is large.
\end{abstract}

\section{Introduction}

The Standard Model of particle physics successfully describes known
collider experiments. With the 10 year old discovery of the heavy
top quark and the more recent identification of the tau lepton the
only missing ingredient of the Standard Model is the elementary Higgs
field. In the minimal Standard Model a weak doublet (hypercharge Y=1)
scalar field is postulated with an \textit{ad hoc} scalar potential
to trigger electroweak symmetry breaking. Three Goldstone Bosons are
eaten up by the $W^{\pm},Z$ and the remaining single CP-even neutral
Higgs scalar has evaded the experimental discovery so far. There is
only a lower bound from LEP2 experiment $M_{H}>114.5$ GeV and an
upper bound from electroweak precision tests, which was raised considerably
to $M_{H}<$251 GeV after the refined measurement of the top quark
mass \cite{top}. Beside the missing experimental discovery, theories
with elementary scalars are burdened with theoretical problems, like
triviality and the most severe gauge hierarchy problem. Elementary
scalars are unstable against radiative corrections and without fine
tuning the Standard Model must be cut off at 1-2 TeV. We need experiments
at the TeV scale to reveal the true nature of electroweak symmetry
breaking.

There are basicly two ways to solve these problems in particle physics,
either impose new symmetries to protect the scalars or elminate elementary
scalars from the theory. The traditional protector is supersymmetry
leading to the Minimal Supersymmetric Standard Model, which is very
attractive considering the radiatively triggered symmetry breaking,
ideal dark matter candidates and successful gauge coupling unification
in supersymmetric Grand Unified Theories. However supersymmetric theories
involve huge parameter space and doubling of all known particles.
None of the predicted new superpartners have been found in any of
the experiments and supersymmetry starts to lose it's appeal. Recently
{}``little'' Higgs models \cite{lh} attracted considerable interest
solving the {}``little hierarchy problem'' allowing to raise the
cutoff of the theory up to 10 TeV without excessive fine tuning. Little
Higgs models realize the old idea that the Higgs is a pseudo Goldstone
boson of some spontaneously broken global symmetry \cite{pshiggs}.
Contrary to supersymmetric models divergent fermion (boson) loops
cancel fermion (boson) loops. Little Higgs models still require large
fine tuning unless they posess custodial symmetry at the price of
highly extended gauge groups.

Scenarios without elementary scalars in four dimemsions are based
on dynamical symmetry breaking mechanism. One possibility is a symmetry
breaking system interacting strongly with the longitudinal weak vector
bosons which has been realised by Dobado \textit{et al.} in the DHT
model \cite{dht} based on a chiral Lagrangian approach. An alternative
description of the strongly interacting symmetry breaking system has
been proposed in the BESS model \cite{bess} through nonlinear realisations.
Top quark condensation has also been suggested for describing the
electroweak symmetry breaking \cite{nambu} leading to several interesting
studies \cite{topc}. Electroweak symmetry breaking caused by the
condensation of a vector field was studied, too \cite{vcm1}. Condensation
of vector bosons in different scenarios was considered in the literature.
Condensation of the weak gauge bosons $W^{\pm}$ were studied by Linde
in extreme dense fermionic matter \cite{linde}, Ambjorn and Olesen
investigated the $W^{\pm}$ condensation in strong external magnetic
field \cite{ambj}. Introduction of new global symmetry and the role
of a chemical potential was studied even for the Electroweak Symmetry
\cite{sannino} .

Extra dimensional models were recently proposed to solve the hierarchy
problem via changing the high energy (small distance) behaviour of
gravity \cite{add,rs}. Concerning symmetry breaking mechanisms new
studies went back to the idea of Manton \cite{manton} and Hosotani
\cite{hoso} in which the Higgs field is the extra component of a
higher dimensional gauge field \cite{higgsgb}. Little Higgs models
were motivated by these ideas though the simplest models \cite{lsth}
are viable in four dimensions. The latest idea is breaking the electroweak
symmetry without a Higgs \cite{higgsl}. In five dimensions boundary
conditions break the original symmetries and the Kaluza Klein modes
of the gauge bosons play the role of the standard Higgs scalar, e.g.
successfully unitarize the weak gauge boson scattering amplitudes.
Higgsless models can be thought as a gravity dual of walking technicolor
models with the advantage of weakly coupled regimes allowing perturbative
calculations \cite{nom}. In the previous examples we have seen that
vector bosons can play a role in electroweak symmetry breaking.

In this chapter we present a recently proposed alternative model of
electroweak symmetry breaking \cite{vcm2}. Consider the usual Lagrangian
of the standard model of electroweak interactions but instead of the
scalar doublet two new matter fields are introduced. One of them is
a Y = 1, T = 1/2 doublet of complex vector fields \begin{equation}
B_{\mu}=\left(\begin{array}{c}
B_{\mu}^{(+)}\\
B_{\mu}^{(0)}\end{array}\right),\label{eq:bdub}\end{equation}
 the other is a real Y = 0, T = 0 vector field \begin{equation}
C_{\mu}.\label{eq:cmu}\end{equation}
 This extends our recent model \cite{vcm1,vcm3} where only the doublet
$B_{\mu}$ was present with the condensation of $B_{\mu}^{(0)}$.
Consequently, we are able to describe a more complete symmetry breaking
and to generate fermion masses from a gauge invariant interaction
Lagrangian while the mass ratio of the new particles does not become
fixed. The key point is the introduction of a mixed $B_{\mu}-C_{\mu}$
condensate together with suitable gauge invariant interactions of
the new matter fields. This leads to nonvanishing standard model particle
masses, as well as B, C particle masses. It turns out that altogether
three condensate emerge but only one combination of theirs is fixed
by the Fermi coupling constant. The model should be considered as
a low energy effective one. Based on recent experience \cite{vcm2,vcm4},
probably it has a few TeV cutoff scale. Its' new particle content
is a charged vector boson pair and three neutral vector bosons. As
is shown, these can be pair produced in $e^{+}e^{-}$ annihilation
and hadron colliders, and at future linear colliders of 500-1500 GeV
and the LHC they can provide a large number of events.

\section{The model}

To build the model, in the Lagrangian of the standard model the interactions
of the scalar doublet are replaced by the gauge invariant Lagrangian

\begin{eqnarray}
L_{BC} & = & -\frac{1}{2}\overline{\left(D_{\mu}B_{\nu}-D_{\nu}B_{\mu}\right)}\left(D^{\mu}B^{\nu}-D^{\nu}B^{\mu}\right)-\nonumber \\
 &  & -\frac{1}{2}\left(\partial_{\mu}C_{\nu}-\partial_{\nu}C_{\mu}\right)\left(\partial^{\mu}C^{\nu}-\partial^{\nu}C^{\mu}\right)-V(B,C),\label{eq:Lagrangian}\end{eqnarray}
 where $D_{\mu}$ is the covariant derivative, $g_{\mu\nu}=+---$,
and for the potential $V(B,C)$ we assume \begin{equation}
V(B,C)=\lambda_{1}\left(\overline{B}_{\nu}B^{\nu}\right)^{2}+\lambda_{2}\left(C_{\nu}C^{\nu}\right)^{2}+\lambda_{3}\overline{B}_{\nu}B^{\nu}C_{\mu}C^{\mu},\label{eq:vbc}\end{equation}
 depending only on B-, C- lengths. Other quartic terms would not change
the argument. $\lambda_{1,2,3}$ are real and from positivity \begin{equation}
\lambda_{1}>0,\quad\lambda_{2}>0,\;\quad4\lambda_{1}\lambda_{2}>\lambda_{3}^{2}.\label{eq:lambdarel}\end{equation}
 Mass terms to be generated are not introduced explicitely in (\ref{eq:vbc}).
Fermion-BC interactions are introduced later on.

To break the gauge symmetry, we assume a nonvanishing mixed condensate
in the vacuum ,\begin{equation}
\left\langle \overline{B}_{\mu}C_{\nu}\right\rangle =g_{\mu\nu}\:\:(0,d),\label{eq:5}\end{equation}
 where the left-hand side could be rotated into (0,d), $d\neq0$,
respecting also electric charge conservation and defining the neutral
and charged components in (\ref{eq:bdub}). By $U_{Y}(1)$ $d$ can
be chosen real. A real $d$ respects also combined TCP and C symmetries.
By TCP-invariance (\ref{eq:5}) equals $\left\langle C_{\nu}\overline{B}_{\mu}\right\rangle $.
It follows from (\ref{eq:5}) that the only nonvanishing mixed condensate
is \begin{equation}
\left\langle B_{1\mu}C_{\nu}\right\rangle =\sqrt{2}g_{\mu\nu}d\label{eq:7}\end{equation}
 with\begin{equation}
B_{\mu}^{(0)}=\frac{{1}}{\sqrt{{2}}}\left(B_{1\mu}+iB_{2\mu}\right),\label{eq:8}\end{equation}
 where $B_{j\mu}$ is real. Once there exists the mixed condensate,
B and C are assumed to condense separately\begin{eqnarray}
\left\langle \overline{B}_{\mu}B_{\nu}\right\rangle  & = & g_{\mu\nu}k_{1},\quad k_{1}\neq0,\label{eq:9}\\
\left\langle C_{\mu}C_{\nu}\right\rangle  & = & g_{\mu\nu}k_{3},\quad k_{3}\neq0.\nonumber \end{eqnarray}
 Once the condensate in (\ref{eq:5}) defines the broken generators,
(\ref{eq:9}) respects the correct electroweak symmetry breaking pattern,
$k_{1}$ must originate from $B_{\mu}^{(0)}$ condensation, \begin{eqnarray}
\left\langle B_{\mu}^{(+)\dagger}B_{\nu}^{(+)}\right\rangle  & = & 0,\nonumber \\
\left\langle B_{\mu}^{(0)\dagger}B_{\nu}^{(0)}\right\rangle  & = & g_{\mu\nu}k_{1}.\label{eq:10}\end{eqnarray}
 (\ref{eq:10}) reproduces the pattern of gauge particle masses \cite{vcm1}.
All the condensates linear in $B_{\mu}^{(+)}$ vanish by charge conservation.
Finally, we assume in advance, that \begin{equation}
\left\langle B_{\mu}^{(0)}B_{\nu}^{(0)}\right\rangle =\left\langle B_{\mu}^{(0)\dagger}B_{\nu}^{(0)\dagger}\right\rangle =g_{\mu\nu}\:\: k_{2}.\label{eq:k2cond}\end{equation}
 The point is that in general $B_{1\mu}$ and $B_{2\mu}$ belong to
different masses, so that $k_{2}\neq0$. $k_{1,2,3}$ are real and
$k_{1}<0$, $k_{3}<0$, as shown by physical particle masses and simple
models. The condensates are of nonperturbative origin caused by the
strong interaction V(B,C). Among them only $k_{1}$ is fixed by contemporary
phenomenology.

\section{Boson and fermion masses}

Mass terms are obtained in the linearized form of $L_{BC}$ via condensates.
The $W^{\pm}$ mass is determined by the total B-condensate, while
the two neutral gauge field combinations are proportional to $B_{\mu}^{(+)\dagger}B_{\nu}^{(+)}$
and $B_{\mu}^{(0)\dagger}B_{\nu}^{(0)}$, respectively. Therefore,
(\ref{eq:10}) yields\begin{equation}
m_{\textrm{photon }}=0,\; m_{W}=\frac{g}{2}\sqrt{-6k_{1}},\; m_{Z}=\frac{g}{2\cos\theta_{W}}\sqrt{-6k_{1}}.\label{eq:mwz}\end{equation}
 Low energy phenomenology gives\begin{equation}
k_{1}=-\left(6\sqrt{2}G_{F}\right)^{-1},\;\left(-6k_{1}\right)^{1/2}=246\textrm{ GeV}.\label{eq:k1}\end{equation}
 $B^{\pm}$ and $B_{2}$ get the following masses\begin{eqnarray}
m_{\pm}^{2} & = & -8\lambda_{1}k_{1}-4\lambda_{3}k_{3},\label{eq:mpm}\\
m_{B_{2}}^{2} & = & -10\lambda_{1}k_{1}+2\lambda_{1}k_{2}-4\lambda_{3}k_{3}=m_{\pm}^{2}+2\lambda_{1}(k_{2}-k_{1}).\nonumber \end{eqnarray}
 For $\lambda_{3},-k_{1},-k_{3}>0$, $\; m_{B_{2}}^{2}>m_{\pm}^{2}>0$
since $k_{2}>k_{1}$. The $B_{1}-C$ sector is slightly more complicated,
here one arrives at the following bilinear combinations in the potential
for $B_{1\mu},C_{\mu}$\begin{equation}
V(B,C)\rightarrow-\frac{m_{1}^{2}}{2}B_{1\mu}B^{1\mu}-\frac{m_{2}^{2}}{2}C_{\nu}C^{\nu}-m_{3}^{2}B_{1\mu}C^{\mu},\label{eq:14}\end{equation}
 with \begin{eqnarray}
-m_{1}^{2} & = & 10\lambda_{1}k_{1}+2\lambda_{1}k_{2}+4\lambda_{3}k_{3}=-m_{B_{2}}^{2}+4\lambda_{1}k_{2},\nonumber \\
-m_{2}^{2} & = & 24\lambda_{2}k_{3}+8\lambda_{3}k_{1},\label{eq:m123}\\
-m_{3}^{2} & = & 4\sqrt{2}\lambda_{3}d.\nonumber \end{eqnarray}
 Here $m_{1}^{2}>0$ being $k_{1}+k_{2}<0$; $m_{2}^{2}>0$ and $m_{3}^{2}\lessgtr0$.
A positive potential in (14) requires\begin{equation}
m_{1}^{2},m_{2}^{2}>0,\;\quad m_{1}^{2}m_{2}^{2}>m_{3}^{4}.\label{eq:16}\end{equation}

(\ref{eq:14}) shows that $B_{1\mu}$ and $C_{\mu}$are nonphysical
fields, the mass eigenstates are defined by\begin{eqnarray}
B_{f\mu} & = & cB_{1\mu}+sC_{\mu},\nonumber \\
C_{f\mu} & = & -sB_{1\mu}+cC_{\mu},\label{eq:mphys}\end{eqnarray}
 where $c=\cos\phi,\; s=\sin\phi,$ $\phi$ denotes the mixing angle
defined by \begin{equation}
\frac{1}{2}\sin2\phi(m_{1}^{2}-m_{2}^{2})=\cos2\phi m_{3}^{2}.\label{eq:18}\end{equation}
 The physical masses are\begin{eqnarray}
m_{B_{f}}^{2} & = & c^{2}m_{1}^{2}+s^{2}m_{2}^{2}+2csm_{3}^{2},\nonumber \\
m_{C_{f}}^{2} & = & s^{2}m_{1}^{1}+c^{2}m_{2}^{2}-2csm_{3}^{2},\label{eq:mbc}\end{eqnarray}
 whence\begin{eqnarray}
2m_{B_{f}}^{2} & = & m_{1}^{2}+m_{2}^{2}+\frac{m_{1}^{2}-m_{2}^{2}}{\cos2\phi}.\label{eq:m12}\\
2m_{C_{f}}^{2} & = & m_{1}^{2}+m_{2}^{2}-\frac{m_{1}^{2}-m_{2}^{2}}{\cos2\phi}\label{eq:m12b}\end{eqnarray}
 For $(m_{1}^{2}-m_{2}^{2})/\cos2\phi>0$ ($<0$) $m_{B_{f}}^{2}>m_{C_{f}}^{2}>0$
($m_{C_{f}}^{2}>m_{B_{f}}^{2}>0$). At vanishing mixing, $m_{3}^{2}=0$,
$B_{1\mu}$ and $C_{\mu}$ become independent having the masses $m_{1}$
and $m_{2}$; taking $k_{2}=0$ and omitting $C_{\mu}$ we recover
the model of Ref.5. $k_{2}$ shifts the real component field masses
from the mass of the imaginary part $B_{2\mu}$.

The particle spectrum of the B-C sector consists of the spin-one $B^{\pm}$
and the three neutral spin-one particles $B_{2},B_{f},C_{f}$. Their
masses are rather weakly restricted. Beside the gauge coupling constants
and $\lambda_{1},\lambda_{2},\lambda_{3}$, the model has three basic
condensates $\left\langle V_{i\mu}V_{i\nu}\right\rangle $, $V_{i\mu}=B_{2\mu},B_{f\mu},C_{f\mu}$.
$k_{1},k_{2},k_{3},d$ condensates are built up from these as follows\begin{eqnarray}
g_{\mu\nu}d & = & \frac{1}{\sqrt{2}}cs\left(\left\langle B_{f\mu}B_{f\nu}\right\rangle -\left\langle C_{f\mu}C_{f\nu}\right\rangle \right),\nonumber \\
g_{\mu\nu}k_{1} & = & \frac{1}{2}\left\{ c^{2}\left\langle B_{f\mu}B_{f\nu}\right\rangle +s^{2}\left\langle C_{f\mu}C_{f\nu}\right\rangle +\left\langle B_{2\mu}B_{2\nu}\right\rangle \right\} ,\nonumber \\
g_{\mu\nu}k_{2} & = & \frac{1}{2}\left\{ c^{2}\left\langle B_{f\mu}B_{f\nu}\right\rangle +s^{2}\left\langle C_{f\mu}C_{f\nu}\right\rangle -\left\langle B_{2\mu}B_{2\nu}\right\rangle \right\} ,\label{eq:k123d}\\
g_{\mu\nu}k_{3} & = & s^{2}\left\langle B_{f\mu}B_{f\nu}\right\rangle +c^{2}\left\langle C_{f\mu}C_{f\nu}\right\rangle .\nonumber \end{eqnarray}
 From (\ref{eq:k123d}) $d$ can be written as \begin{equation}
2\sqrt{2}\cot2\phi\: d=k_{1}+k_{2}-k_{3}.\label{eq:d-k}\end{equation}

Turning to the dynamical fermion mass generation, we add to the gauge
vector and matter vector field Lagrangians, the usual fermion-gauge
vector Lagrangian, as well as a new gauge invariant piece responsible
for the fermion-matter vector field interactions and in usual notation
this is (for quarks)\begin{eqnarray}
g_{ij}^{u}\overline{\psi}_{iL}u_{jR}B_{\nu}^{C}C^{\nu}+g_{ij}^{d}\overline{\psi}_{iL}d_{jR}B_{\nu}c^{\nu}+h.c.,\label{eq:mfermion}\\
\psi_{iL}=\left(\begin{array}{c}
u_{i}\\
d_{i}\end{array}\right)_{L},\quad B_{\nu}^{C}=\left(\begin{array}{c}
B_{\nu}^{(0)\dagger}\\
-B_{\nu}^{(+)\dagger}\end{array}\right).\nonumber \end{eqnarray}

Clearly the mixed condensate provides fermion masses and also the
Kobayashi-Maskawa description is unchanged. A typical fermion mass
is\begin{equation}
m_{f}=-4g_{f}d\label{eq:mfermd}\end{equation}
 and only $g_{f}d$ becomes fixed but $m_{f1}/m_{f2}=g_{f1}/g_{f2}$
as usual. If $d$ is about $k_{1}\simeq G_{F}^{-1}$, then $g_{f}$
is a factor of $G_{F}^{1/2}$ weaker than the approximate standard
model value $G_{F}^{1/2}$.

\section{Interactions of the new vector bosons}

The new physical fields $B_{f\mu},C_{f\mu},B_{2\mu},B_{\mu}^{+}$
have various interactions. Four-boson self-interactions can be read
off from the V(B,C) potential (\ref{eq:vbc}), the coupling are all
proportional to some unknown $\lambda_{i}$.

The new vector bosons interact with the standard fermions via the
Yukawa interactions (\ref{eq:mfermion}), but the coupling strength
is expected to be weaker than the Standard Model Higgs-fermion couplings.

The most important interactions of the new particles from the point
of view of phenomenology are the one with the gauge bosons. The source
of these interaction is the $B_{\mu}$ covariant derivative terms
in (\ref{eq:Lagrangian}). The three particle interaction all have
a derivative coupling, before the B-C mass diagonalization they are

\begin{eqnarray}
L_{int}^{(3)}= & -ie\left(\partial_{\mu}B_{\nu}^{(-)}-\partial_{\nu}B_{\mu}^{(-)}B^{(+)\nu}\right)A^{\mu}+\nonumber \\
 & +ie\cot2\theta_{w}\left(\partial_{\mu}B_{\nu}^{(-)}-\partial_{\nu}B_{\mu}^{(-)}\right)B^{(+)\nu}Z^{\mu}-\nonumber \\
 & -i\frac{\sqrt{2}}{2}\frac{e}{\sin\theta_{w}}\left(\partial_{\mu}B_{\nu}^{(-)}-\partial_{\nu}B_{\mu}^{(-)}\right)B^{(0)+\nu}W^{+\mu}-\nonumber \\
 & -i\frac{\sqrt{2}}{2}\frac{e}{\sin\theta_{w}}\left(\partial_{\mu}B_{\nu}^{(0)+}-\partial_{\nu}B_{\mu}^{(0)+}\right)B^{(+)\nu}W^{-\mu}-\nonumber \\
 & -i\frac{e}{\sin2\theta_{w}}\left(\partial_{\mu}B_{\nu}^{(0)+}-\partial_{\nu}B_{\mu}^{(0)+}\right)B^{(0)\nu}Z^{\mu} & +h.c.\label{int}\end{eqnarray}
In terms of the physical fields the neutral boson interactions change
considerably, as an example we give the $Z-B_{f}-B_{2}$ coupling,

\begin{equation}
L_{int,2}=\frac{g}{2\cos\theta_{W}}\cos\phi\cdot Z_{\mu}\left[B_{f\nu}(\partial^{\mu}B_{2}^{\nu}-\partial^{\nu}B_{2}^{\mu})-B_{2\nu}(\partial^{\mu}B_{f}^{\nu}-\partial^{\nu}B_{f}^{\mu})\right].\label{eq:Lint}\end{equation}

There exist several $B_{i}B_{j}VV-$ type four particle couplings
with gauge bosons with $V=\gamma,W^{\pm},Z$ and $B_{i\mu}=B_{f\mu},C_{f\mu},B_{2\mu},B_{\mu}^{+}$
, the coupling strength is $\sim g^{2}$ multiplied with mixing angles
(e.g. $\cos\theta_{W},\sin\phi$). We remind here that there is no
mixing between the gauge and new vector bosons. As an example we give
the $VVB^{+}B^{-}$couplings\begin{eqnarray}
L_{int,+}^{(4)} & = & \left(g^{\mu\alpha}g^{\nu\beta}-g^{\mu\nu}g^{\alpha\beta}\right)\left[e^{2}A_{\nu}A_{\mu}B_{\alpha}^{(-)}B_{\beta}^{(+)}+\frac{g^{2}}{2}W_{\mu}^{(-)}W_{\mu}^{(+)}B_{\alpha}^{(-)}B_{\beta}^{(+)}+\right.\label{eq:L4+}\\
 &  & \left.+e^{2}\cot^{2}2\theta_{W}Z_{\mu}Z_{\nu}B_{\alpha}^{(-)}B_{\beta}^{(+)}-e^{2}\cot2\theta_{W}\left(A_{\mu}Z_{\nu}B_{\alpha}^{(-)}B_{\beta}^{(+)}+Z_{\mu}A_{\nu}B_{\alpha}^{(-)}B_{\beta}^{(+)}\right)\right]\nonumber \end{eqnarray}

All the interactions have even number of new $B,C$ particles, the
single production of the new particles are not allowed in this model.
The new particles can only decay to each other determined by their
mass relation making them difficult to discover in particle physics
experiment and the lightest one is stable. At the same time if the
lightest new vector boson is neutral then it will be an ideal candidate
for a selfinteracting dark matter. Generally all the couplings are
weaker than the relevant couplings of one standard Higgs boson and
the most important ones are the three boson couplings in (\ref{int},\ref{eq:Lint}).
In what follows we investigate the implications of the interactions.

\section{Unitarity constraints}

In this section we apply tree-level partial wave unitarity to two-body
scatterings of longitudinal gauge and B,C bosons following the reasoning
of \cite{lee}, where perturbative unitarity has been employed to
constrain the Standard Model Higgs mass. With the only experimental
input of the muon decay constant $G_{F}$ the perturbative upper bound
of appoximately 1 TeV emerged for the Standard Model Higgs mass. Perturbative
unitarity is a powerful tool, it can be used to build up the bosonic
sector of the Standard Model and it was essential to build higssless
models of electroweak symmetry breaking in extra dimensional field
theories. Perturbative unitarity shows that the scale of the model
is approximately 2.5 TeV and the new particle masses are bounded from
below, the lower bound increasing with a growing $\Lambda$ \cite{vcm4}. 

In the vector condensate model there exist many elastic $VV\rightarrow B_{i}B_{j}$
and $B_{i}V\rightarrow B_{i}V$ type processes, with $V=\gamma,W^{\pm},Z$
and $B_{i}=B_{f},C_{f},B_{2},B^{+}$. We consider these processes
for longitudinally polarized external particles, as these are the
most dangerous one to ruin the high energy behaviour, in the high
energy limit \begin{equation}
\epsilon_{\mu}(k)\sim\frac{k_{\mu}}{M},\label{eq:long}\end{equation}
 where $M$ is the mass of the outgoing/incoming vector particle.
We calculate then the $J=0$ partial-wave amplitudes, $a_{0}$, from
contact and one-particle exchange graphs. Unitarity requires $\left|Re\: a_{0}\right|\leq\frac{1}{2}.$

\begin{center}%
\begin{figure}[H]
\includegraphics[%
  scale=0.65]{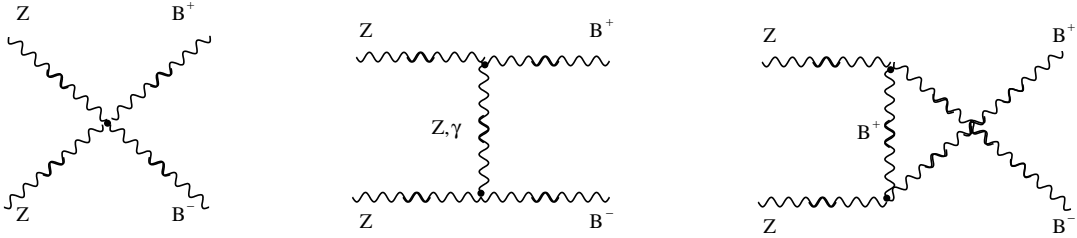}

\caption{Tree level Feynman\label{cap:bbzz} graphs for $B^{+}B^{-}\rightarrow B^{+}B^{-}$
.}
\end{figure}
\end{center}

The scattering processes provide vastly different constraints on the
parameters of the theory. First choose a process with known coupling
constants. In the process $ZZ\rightarrow B^{+}B^{-}$ all the couplings
are proportional to the weak coupling constant ($e\,\cot2\theta_{W}$).
There are three graphs shown in Fig \ref{cap:bbzz}. contributing
to the elastic scattering, the contact graph with the four particle
vertex and t- and u-channel graphs. In the $s\gg m_{+},m_{Z}$ limit
applying (\ref{eq:long}) the contributions are the following ($m_{+}=m_{B^{+}})$\begin{eqnarray*}
T_{c} & = & e^{2}\frac{\cot^{2}2\theta_{W}}{m_{Z}^{2}m_{+}^{2}}\left(\frac{s^{2}}{2}-\frac{t^{2}}{4}-\frac{u^{2}}{4}\right),\\
T_{t} & = & e^{2}\frac{\cot^{2}2\theta_{W}}{m_{Z}^{2}m_{+}^{2}}\left(\frac{1}{4}t(s-u)\right),\\
T_{u} & = & e^{2}\frac{\cot^{2}2\theta_{W}}{m_{Z}^{2}m_{+}^{2}}\left(\frac{1}{4}u(s-t)\right).\end{eqnarray*}
The sum of the three amplitude growing with the energy vanishes\[
T_{c}+T_{t}+T_{u}=0,\]
reflecting that the $B^{+}B^{-}Z$ interactions can in principle originate
from a renormalizable interactions and the high energy behaviour of
the process is modest. A detailed calculation with the general polarizations
give for the s-wave amplitude\[
\left|a_{0}\right|=\frac{e^{2}\cot^{2}2\theta_{W}}{32\pi}\frac{m_{Z}^{2}}{m_{+}^{2}}+{\cal O}(1/s),\]
giving $m_{+}\geq3$GeV and similar weak bounds emerge from other
elastic $BV\rightarrow BV$ and $BB\rightarrow VV$ scatterings.

\begin{center}%
\begin{figure}[H]
\includegraphics[%
  scale=0.65]{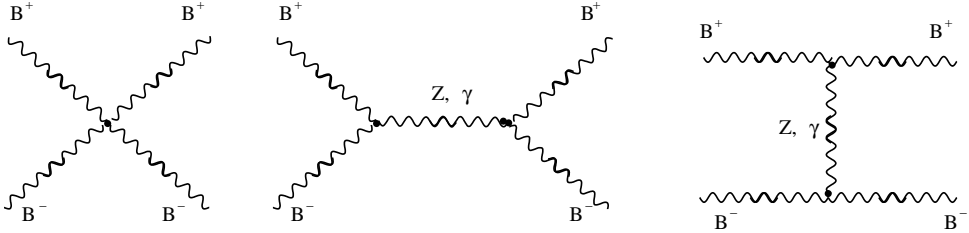}

\caption{Tree level Feynman\label{cap:bbbb} graphs for $B^{+}B^{-}\rightarrow B^{+}B^{-}$
.}
\end{figure}
 \end{center}

Strong bounds emerge from the really non-renomalizable sector of the
theory, the quartic B
 interactions. Taking $\lambda_{3}k_{3}$ negligible in (\ref{eq:mpm})
$\lambda_{1}$ is proportional to $m_{+}^{2}G_{F}$. Consider the
tree level process $B^{+}B^{-}\rightarrow B^{+}B^{-}$ via the contact
graph and the $Z,\gamma$ exchange graphs in the s and t channels
(Fig \ref{cap:bbbb}.). In the limit $s\gg m_{+}^{2},m_{Z}^{2}$ only
the contribution of the contact graph is important\[
T_{c}=\lambda_{1}\frac{1}{2m_{+}^{4}}\left(t^{2}+u^{2}\right),\]
yielding for the s-wave amplitude\[
\left|a_{0}\right|=\frac{\sqrt{2}}{64\pi}G_{F}\frac{s^{2}}{m_{+}^{2}}.\]
The unitarity constraint $\left|Re\: a_{0}\right|\leq\frac{1}{2}$
along with the assumptions that the masses in the model must be smaller
than the cutoff provide the scale of the model $\Lambda\leq2.5$TeV.
The bound is similar also for the $B^{\pm}B^{\pm}\rightarrow B_{2}B_{2}$
scattering. In case of vanishing mixing between $B_{1}$ and $C$
a rough interpretation of $k_{1}$ with a cutoff free propagator shows
similar bounds $\Lambda\leq2-2.6$ TeV depending on the interpretation
of $k_{1}$ \cite{vcm0}. We conclude that perturbative unitarity
estimates suggest that the scale of the model is in the range 2-3
TeV, where a (more) fundamental description takes the role of the
Vector Condesate Model.

\section{Production in $e^{+}e^{-}$colliders}

Direct production of $B_{f}B_{2}$ pairs can be studied in high energy
$e^{+}e^{-}$ colliders, $e^{+}e^{-}\rightarrow Z^{*}\rightarrow B_{f}B_{2}$.
Assume in (\ref{eq:mfermd}) $g_{e^{-}}$ is very small, then the
direct $e^{+}e^{-}\rightarrow B_{f}B_{2}$ can be neglected. From
(\ref{eq:Lint}) we have for the total cross section 

\begin{figure}
\includegraphics[%
  scale=0.5,
  angle=270]{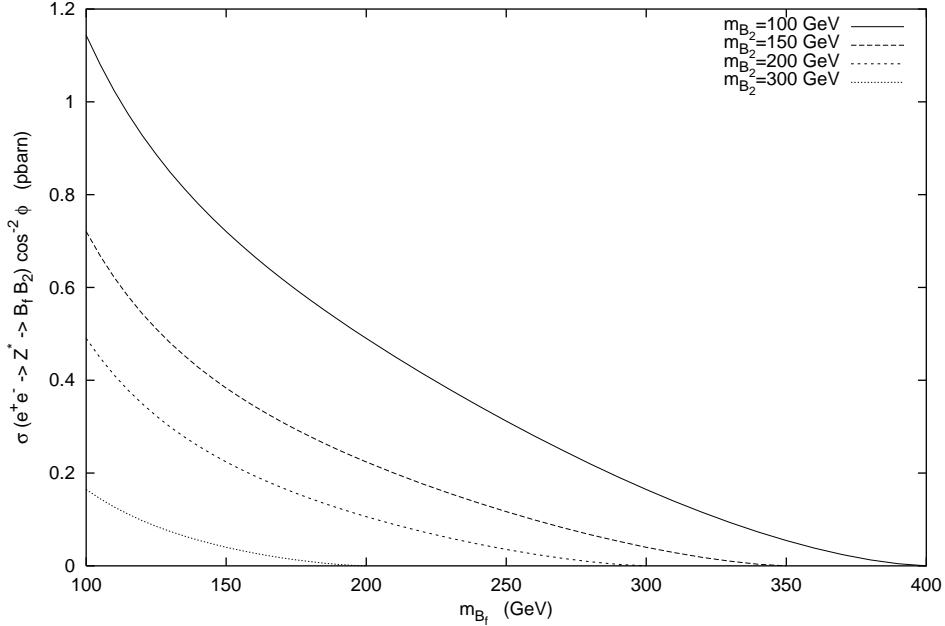}

\caption{$\cos^{-2}\phi\;\sigma\left(e^{+}e^{-}\rightarrow B_{f}B_{2}\right)$
vs. $m_{B_{f}}$at $\sqrt{s}=500$ GeV and various $m_{B_{2}}.$}
\end{figure}
 \[
\sigma(e^{+}e^{-}\rightarrow Z^{*}\rightarrow B_{f}B_{2})=\frac{g^{4}\cos^{2}\phi}{3\cdot4096\cos^{4}\theta_{W}}\frac{1+(4\sin^{2}\theta_{W}-1)^{2}}{m_{B_{2}}^{2}m_{B_{f}}^{2}s^{2}(s-M_{Z}^{2})^{2}}\left(s-(m_{B_{2}}+m_{B_{f}})^{2}\right)^{3/2}\cdot\]

\begin{equation}
\cdot\left(s-(m_{B_{f}}-m_{B_{2}})^{2}\right)^{3/2}\left(2s(m_{B_{2}}^{2}+m_{B_{f}}^{2})+m_{B_{f}}^{4}+m_{B_{2}}^{4}+10m_{B_{2}}^{2}m_{B_{f}}^{2}\right).\label{eq:sigma}\end{equation}

\begin{figure}
\includegraphics[%
  scale=0.5,
  angle=270]{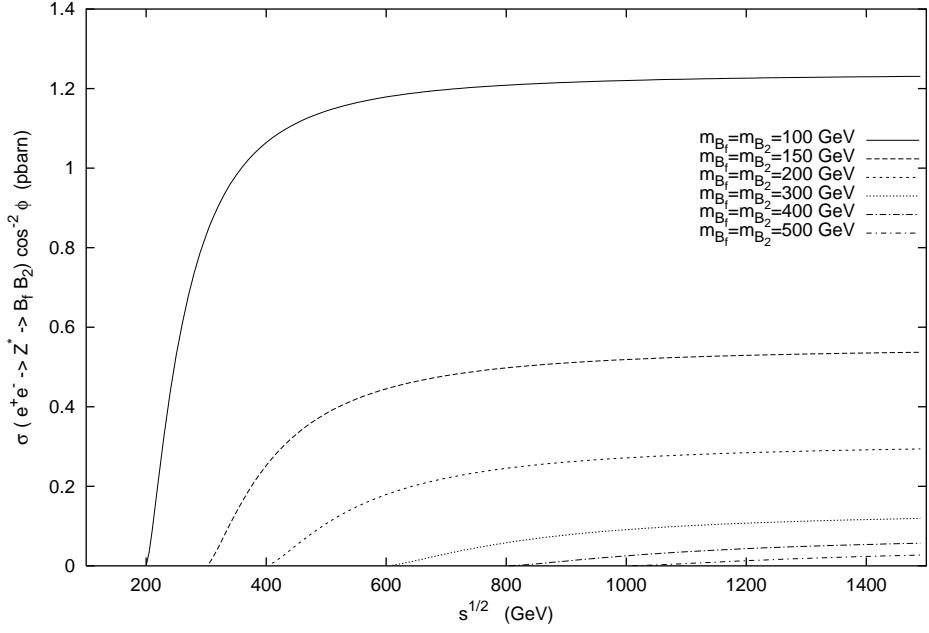}

\caption{$\cos^{-2}\phi\;\sigma\left(e^{+}e^{-}\rightarrow B_{f}B_{2}\right)$
vs. $\sqrt{s}$ at various $m_{B_{f}}=m_{B_{2}}$.}
\end{figure}

At asymptotic energies $\sigma$ is proportional to $1/m_{B_{2}}^{2}+1/m_{B_{f}}^{2}$.
The mass and energy dependences of $\sigma$ are shown in Figs. 3,4.
For example at $\sqrt{s}$= 500 GeV and with an integrated luminosity
of 10 fb$^{-1}$ 5700, 1900, 530 $B_{f}B_{2}$ pairs are expected
for $m_{B_{f}}=m_{B_{2}}$= 100, 150, 200 GeV and $\cos^{2}\phi=1/2$.
At $\sqrt{s}$ = 1.5 TeV a higher mass range can be tested, for $\cos^{2}\phi=1/2$,
a luminosity of 100 fb$^{-1}$ we get the large event numbers 62200,
14500, 5900, 1900, 530 for $m_{B_{f}}=m_{B_{2}}$= 100, 200, 300,
400, 500 GeV. One can show that the $B^{+}B^{-}$ production is a
factor of $\cos^{2}2\theta_{W}$ smaller than (\ref{eq:sigma}) at
equal masses and $\cos^{2}\phi=1$.

Though we expect a large number of events at the next generation of
electron positron colliders the identification of the new particles
will be a difficult task. The lightest of $B,C$ is stable and if
it is a neutral particle it is ideal dark matter candidate then the
same problem expected concerning the detection as in the case of other
dark matter particles. The lightest neutral particle only turns up
in the missing energy channels. When a heavier new particle is produced
it can decay subsequently into a lighter new particle emitting a single
gauge boson or a pair of leptons, the appearance of the standard decay
products at a misplaced vertex together with missing energy will trigger
the hopeful discovery of the vector bosons. The charged $B^{+}$-s
can be identified fairly easier via charged tracks and missing energy
in the calorimeters.

\section{New particles at hadron colliders}

In this section we study the production of B--particles at the most
energetic hadron colliders, ideal for discoveries, Tevatron and LHC.
We show that producing heavy B(C)-particles at LHC is very favourable
having a large cross section while at the Tevatron energy the production
cross section cannot exceed (0.01-0.02) fb which is far below the
discovery limit. 

Since fermions are coupled very weakly to B(C)-pairs in the vector
condensate model, producing B,C-pairs is expected to be more considerable
from virtual $\gamma$ and Z exchanges, that is we consider the Drell-Yan
mechanism \cite{dy}, $p^{(}\overline{p}^{)}\rightarrow B\overline{B}+X$
via quark-antiquark annihilation. 

The Drell-Yan cross section for the above hadronic collisions can
be written as \cite{dy,6} \begin{eqnarray}
\sigma(p^{(}\overline{p}^{)}\rightarrow B\overline{B}+X)=\int_{\tau_{0}}^{1}\, d\tau\int_{\tau}^{1}\,\frac{dx}{2x}\sum_{i}\sigma(q_{i}\overline{q}_{i}\rightarrow B\overline{B})\cdot\nonumber \\
\left(f_{i}^{1}(x,\hat{s})f_{\bar{i}}^{2}(\tau/x,\hat{s})+f_{\bar{i}}^{1}(x,\hat{s})f_{i}^{2}(\tau/x,\hat{s})\right),\label{eq:sigmahadron}\end{eqnarray}
 where $x$ and $\tau/x$ are the parton momentum fractions, $\hat{s}=\tau s$
is the square of the centre of mass energy of $q_{i}\bar{q}_{i}$,
s is the same for the hadronic initial state, $f_{i}^{1}(x,\hat{s})$
means the number distribution of $i$ quarks in hadron 1 at the scale
$\hat{s}$ and the sum runs over the quark flavours u,d,s,c. In the
computation the MRS (G) fit program \cite{mrs} was used for the parton
distributions. 

The angle integrated, colour averaged annihilation cross section $\sigma(q_{i}\overline{q}_{i}\rightarrow B^{+}B^{-})$
is calculated to lowest order in the gauge couplings, and QCD corrections
are neglected. We hope this approximation shows the order of magnitude
of the cross section. We give the result of the charged final state
as there is no unknown mixing angle in the result and the identification
of $B^{\pm}$ seems less dificult than for neutral pairs. The $B^{+}B^{-}$
pairs appear via $\gamma+Z$ exchange, the relevant interactions are
in (\ref{int}). At the $q_{j}\overline{q}_{j}Z$-vertex the usual
coupling $ig\gamma_{\mu}(g_{Vj}+g_{Aj}\gamma_{5})$ acts, here 

\begin{eqnarray}
\left.\begin{array}{cc}
g_{Vj}= & \frac{1}{2}-\frac{4}{3}\sin^{2}\theta_{W},\\
g_{Aj}= & \frac{1}{2}\end{array}\right\} j=u,c\nonumber \\
\\\left.\begin{array}{cc}
g_{Vj}= & -\frac{1}{2}+\frac{2}{3}\sin^{2}\theta_{W},\\
g_{Aj}= & -\frac{1}{2}\end{array}\right\} j=d,s\,\,.\nonumber \end{eqnarray}

From (\ref{int}) we get for $B^{+}B^{-}$ final states \cite{vcmhad} 

\begin{eqnarray}
\sigma\left(q_{i}\overline{q}_{i}\rightarrow B^{+}B^{-}\right) & = & \left[(g_{Vi}^{2}+g_{Ai}^{2})cos^{2}2\theta_{W}+2Q_{q_{i}}g_{Vi}sin^{2}2\theta_{W}cos2\theta_{W}+4Q_{q_{i}}^{2}sin^{4}2\theta_{W}\right]\nonumber \\
 &  & \frac{1}{3}\frac{1}{256\pi}\left(\frac{g}{cos\theta_{W}}\right)^{4}\left(1-\frac{4m_{+}^{2}}{\hat{s}}\right)^{3/2}\frac{\hat{s}+8m_{+}^{2}}{4m_{+}^{4}}\label{eq:sigmabb}\end{eqnarray}

\begin{center}%
\begin{figure}
\includegraphics{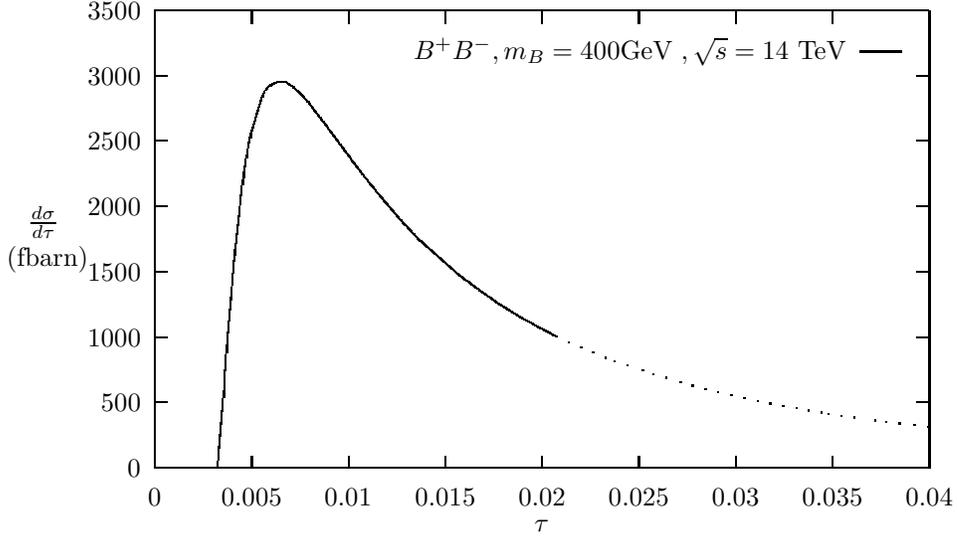}

\caption{\label{cap:ppfig1}The differential $\tau$-distribution of $B^{+}B^{-}$
pairs at LHC, $m_{+}=400$ GeV.}
\end{figure}
 \end{center}

This is decreasing at high, increasing $m_{+}$ and for $\hat{s}\gg4m_{+}^{2}$
it is proportional to $\hat{s}/m_{+}^{4}$ reflecting that the Lagrangian
(\ref{eq:Lint}) is coming from the nonrenormalizable, effective model.
The individual terms are due to $Z$ exchange, $\gamma-Z$ interference
and $\gamma$ exchange. The production of neutral particle pairs $B_{f}B_{2}$
($C_{f}B_{2}$) is only realized via the Z exchange channel includes
an undetermined mixing angle factor, $\cos^{2}\phi$ ($\sin^{2}\phi$)
and is slighty more involved because of two mass parameters. 

\begin{center}%
\begin{figure}
\includegraphics{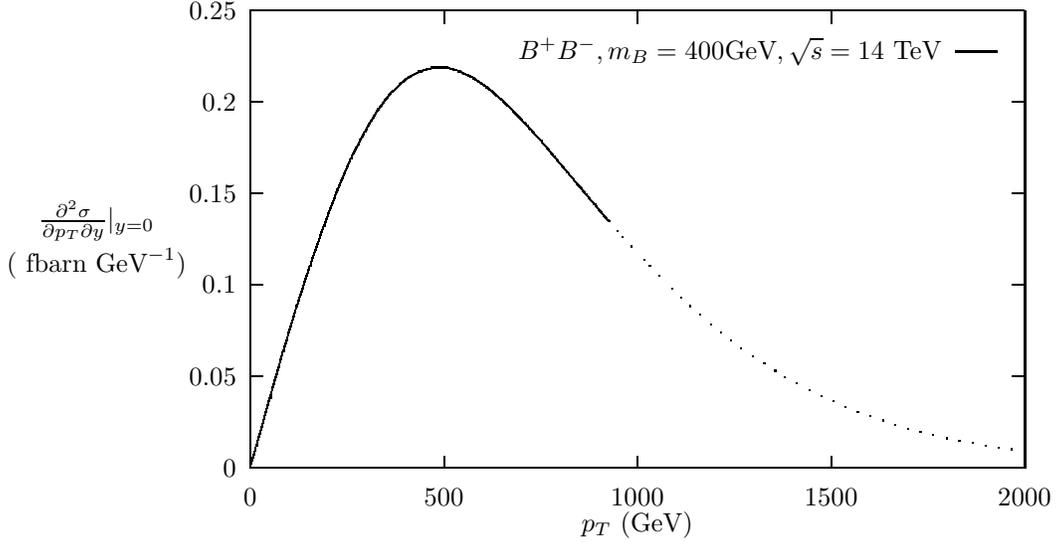}

\caption{\label{cap:ppfig2}$\frac{\partial^{2}\sigma}{\partial p_{T}\partial y}|_{y=0}$
at LHC, $m_{+}=400$ GeV. $p_{T}$ denotes the transverse momentum
of $B^{+}$ and $y$ is the rapidity of $B^{+}B^{-}$. }
\end{figure}
\end{center}

We have calculated various distributions of $B^{+}B^{-}$ pairs for
$p\overline{p}$ collisions at $\sqrt{s}=1.8$ TeV and for $pp$ collisions
at $\sqrt{s}=14$ TeV, assuming $m_{+}=400,500,600$ GeV. Typically,
at the Tevatron no notable result can be presented, however, increasing
the energy to LHC, we get sizable cross sections. Also the yield of
$B^{+}B^{-}$ is larger than that of neutral pairs $B_{f}B_{2},C_{f}B_{2}$.
As an example we show in Fig.\ref{cap:ppfig1}. the $\tau$-distribution
of $B^{+}B^{-}$ pairs at LHC, $m_{+}=400$ GeV. $\frac{d\sigma}{d\tau}$
is sharply peaked after threshold ($4m_{+}^{2}/s$) and decreases
for higher invariant masses of $B^{+}B^{-}.$ Calculating the total
cross section the dotted part of $d\sigma/d\tau$ was not integrated
corresponding to a cutoff $\sqrt{\hat{s}}=2$ TeV at the parton level
($\tau\leq0.02$). Fig. \ref{cap:ppfig2}. shows $\frac{\partial^{2}\sigma}{\partial p_{T}\partial y}|_{y=0}$
as the function of the transverse momentum $p_{T}$ of $B^{+}$ at
vanishing rapidity y of $B^{+}B^{-}$ for LHC, $m_{+}=400$ GeV. The
start of the dotted curve correspond to $\sqrt{\hat{s}}=2$ TeV. 

For the total cross section (\ref{eq:sigmahadron}) we obtain \begin{eqnarray}
 & \sigma_{\textrm{Tev}} & =0.020\hbox{\, fb\quad for}\qquad B^{+}B^{-},m_{+}=400\hbox{GeV},\nonumber \\
 & \sigma_{LHC} & =33.0;8.5;2.4\,{\textrm{fb }}\hbox{\quad for}\qquad B^{+}B^{-},m_{+}=0.4;0.5;0.6\hbox{TeV}.\end{eqnarray}

From Fig. \ref{cap:ppfig1}. we can immediately read off the cutoff
dependence. For instance, for $m_{+}=400$ GeV, $\sigma_{LHC}(B^{+}B^{-})=20.5(36.8)$
fb at a cutoff 1.5(2.5) TeV. At an expected integrated luminosity
of $10^{5}\textrm{pb}^{-1}$ one gets about 3300 $B^{+}B^{-}$ pairs
of $m_{+}=0.4$ TeV at LHC per annum at a cutoff 2 TeV. The detection
of the new particles similar to the case of electron-positron colliders,
though it is even more difficult, charged tracks can be searched for,
but the high luminosity provides large number of events.

In this section, we have shown that heavy B--particle pairs have a
large inclusive cross section due to $q\overline{q}$ annihilation
at the LHC in the few hundred GeV mass range making the detection
of $B^{+}$ or neutral partners at LHC possible.

\section{Conclusion}

In this chapter a low energy dynamical symmetry breaking model of
electroweak interactions based on matter vector field condensation
is introduced. Mass generation is arranged starting from gauge invariant
Lagrangians, fermion masses are also coming from the vector condensates.
New particles are all spin-one states, one charged pair and three
neutral particles having various interactions. The masses of the new
particles are assumed to generated also by condensation resulting
a nontrivial mixing among the neutral components. The production of
the new particles in electron-positron collider was studied and a
large number of event is expected at future colliders with the center
of mass energy 500-1000 GeV. Tevatron cannot produce enough new particles
for observation, but the yield at the LHC expected to be well above
the discovery limit for new particles with masses of a few hundred
GeV. At present there is only a $\sim$45 GeV direct lower bound from
the invisible Z width \cite{vcm0}. Further constraints can arise
from the electroweak precision tests of the Standard Model. In the
model \cite{vcm1} proposed earlier the precison S,T,U parameters
\cite{peskin} were calculated \cite{vcm1}, and found to be in agreement
with the latest experimental data. The parameter space of the model
presented in this chapter is larger than that of the one in \cite{vcm1},
therefore, we expect that the positive result of the S,T parameter
analysis can be maintained. It would be interesting and exciting to
investigate in details the one loop radiative corrections in our model
and to confront with available experiments.

\end{document}